\documentclass[twocolumn,showpacs]{revtex4}
\usepackage{graphicx}
\begin{document}

\title{Connecting neutron star observations to the high density equation of state of quasi-particle model}
\author{Yan Yan$^{1}$, Jing Cao$^{1}$,  Xin-Lian Luo$^{2}$, Wei-Min Sun$^{1,3}$ and Hongshi Zong$^{1,3}$\footnote{Email:zonghs@chenwang.nju.edu.cn}}
\address{$^{1}$Department of Physics, Nanjing University, Nanjing, 210093, China}
\address{$^{2}$Department of Astronomy, Nanjing University, Nanjing
210093,China}
\address{$^{3}$ Joint Center for Particle, Nuclear Physics and Cosmology, Nanjing 210093, China}
\begin{abstract}
The observation of $1.97\pm0.04$ solar-mass neutron-like star gives
constraint on the equation of state (EOS) of cold, condensed matter.
In this paper, the EOS for both pure quark star and
hybrid star with a quark core described by quasi-particle model are
considered. The parameters of quasi-particle model which affect the
mass of both quark star and hybrid star can be constrained by the
observation.


\end{abstract}

\maketitle

 \maketitle
\section{Introduction}
It is well known that the equation of state (EOS) for cold and dense
strongly interacting matter plays a key role in the study of the
neutron stars in astrophysics \cite{Ozel,Alford}. The neutron star
are presumed to contain the densest matter in the cosmos, which
provide a natural laboratory for cold, condensed matter. The recent
observation \cite{2solarmass} of the pulsar PSR J1614-2230 with a
mass of $1.97\pm0.04 M_{\bigodot}$ ($M_{\bigodot}$ denotes the
mass of the sun) gives a strong constraint on the
EOS of strongly interacting matter.

Just as the discovery of neutron led to the idea of neutron star,
the conception of quark naturally stimulated many physicists and
astrophysicists to suggest that neutron-like star may be quark or
hybrid star (Quark or hybrid star is composed, in whole or in part,
of quark matter) \cite{TH-401,TH-412,N,K,K1,K2}. At present, it is
still not possible to get a reliable EOS of quark matter from the first
principle of Quantum Chromodynamics (QCD). In this case, a commonly
used approach in the study of high density quark matter is to give
up on first principle calculations and resort to phenomenological
QCD models, such as MIT bag model \cite{Weisskopf,Weber,Soff,Paris},
NJL model \cite{Rehberg,Greiner,Ruster,Menezes,Jiang}, perturbative
QCD model \cite{Freedman,Baluni,Fraga,Farhi} and quasi-particle
model \cite{Satz,Peshier,Yang,Cao}. Among these models, the
quasi-particle quark-gluon plasma model with a few fitting
parameters, has been widely used to reproduce the properties of the
QCD plasma at finite temperature and zero chemical potential
\cite{Soff,14,11,15,12,22,13,31}. However, due to lack of experimental
data on dense strongly interacting matter at zero temperature and
finite chemical potential, the
phenomenological parameters of quasi-particle model at zero
temperature and finite chemical potential has some uncertainties. The main purpose of this
paper is try to constrain the parameters of the quasi-particle model
at finite chemical potential based on the most recent astronomical
observation \cite{2solarmass}.

This paper is organized as follows. In Section II, we introduce the
quasi-particle model and its parameters. In Section III, the
structure of neutron-like star is discussed. The relation between
the parameters of quasi-particle model and mass-radii relation of
neutron-like star is also illustrated. Through comparison of
theoretical results with astrophysical observation \cite{2solarmass},
one can give a strong constraint on the parameters of quasi-particle
model used in the present work. Finally, a summary is given in
Section IV.

\section{The Quasi-Particle Model and its Parameters}

Recently, by means of functional integral formalism a
model-independent formula for calculating the EOS of QCD at finite
chemical potential and zero temperature was proposed in Ref.
\cite{ZongSun}. By using this formula the authors in Ref. \cite{ZAM}
obtained an explicit analytic expression for the EOS using the
quasi-particle model. In this paper, we shall use the EOS of this
model to study the structure of the neutron-like star. The EOS in
Ref. \cite{ZAM} is obtained by using path integral method with a
effective quark propagator. This quark propagator has the form of a
free quark propagator with a density dependent effective mass.

According to Ref. \cite{ZongSun}, the quark number density reads
\begin{equation}
\rho(\mu)=-N_{c}N_{f}\int\frac{d^{4}p}{(2\pi)^{4}}tr\{G[\mu](p)\gamma_{4}\},
\end{equation}
where $G[\mu](p)$ is the full quark propagator at finite chemical
potential $\mu$, $N_{c}$ and $N_{f}$ denote the number of colors and
of flavors, respectively, and the trace operation is over Dirac
indices. According to the above formula, in order to calculate the
quark number density, one should know the full quark propagator at
finite $\mu$ in advance. Unfortunately, at present no one knows how
to calculate the full quark propagator from first principles of QCD.
So, one has to choose some effective model quark propagator as the approximation of the
full quark propagator. In this paper we choose the following model
quark propagator proposed in Ref. \cite{ZAM}
\begin{equation}
G^{-1}[\mu](p)=i\gamma\cdot\widetilde p+m(\mu),
\end{equation}
where  ${\widetilde p}\equiv ({\vec p}, p_4+i\mu)$ and the Euclidean
gamma matrices satisfy the algebra
$\{\gamma_{\mu},\gamma_{\nu}\}=2\delta_{\mu\nu}$. Here, following
Refs. \cite{PRC72025809, PRD77034004}, we choose the effective quark
mass $m(\mu)$ and the effective coupling constant $g^{2}(\mu)$ to
be:
\begin{equation}
m^{2}(\mu)=\frac{\mu^{2}g^{2}(\mu)}{3\pi^{2}},~~g^{2}(\mu)=\frac{16\pi^{2}}{9\ln(a(\mu+c))^{2}},
\end{equation}
where $a$, $c$ are phenomenological parameters of quasi-particle
model. In this work, the current mass of u,d,s quarks is taken to be
zero. It could satisfy chemical equilibrium and neutrality
naturally. Then we can use the contour integration method to obtain
the quark number density
\begin{equation}
\rho(\mu)=\frac{N_{c}N_{f}}{3\pi^{2}}(\mu^{2}-m^{2}(\mu))^{3/2}\theta(\mu-m(\mu)).
\end{equation}
Due to the step function in the right-hand side of Eq. (4), it can
be found that the quark number density vanishes when $\mu$ is below
a critical value $\mu_0$ (In this work, the value of $\mu_0$ depends on the parameters $a$ and $c$). Namely, $\mu=\mu_0$ is a singularity which separates
two regions with different quark number densities. This behavior
agrees qualitatively with the general conclusion of Ref.
\cite{Halasz}. Now let us turn to the calculation of the EOS.
According to Refs. \cite{JPG342655,PRD78054001}, the EOS of QCD at
$T=0$ reads
\begin{equation}
P(\mu)=P(\mu)|_{\mu=0}+\int_{0}^{\mu} d\mu^{\prime}
\rho(\mu^{\prime}),
\end{equation}
where $P(\mu)|_{\mu=0}$ is vacuum pressure which represents the
pressure density at $\mu=0$. In this paper, we take
$P(\mu)|_{\mu=0}\equiv -B$, where $B$ is a phenomenological
parameter in our present work (at present one cannot calculate its
numerical value reliably from first principles of QCD). Here it
should be noted that Eq. (5) is a model independent formula. This formula
shows that the pressure density $P(\mu)$ at finite
$\mu$ and zero temperature is totally determined by the quark number
density $\rho(\mu)$ (up to a constant term $P(\mu)|_{\mu=0}$). Therefore,
the quark number density has the correct behavior required by QCD
implies that the pressure $P(\mu)$ also has the correct behavior
required by QCD at finite chemical potential and zero temperature.

Substituting Eq. (4) into Eq. (5), we obtain
\begin{equation}
P(\mu)=-B+\frac{3}{\pi^{2}}\int_{0}^{\mu}d\mu^{\prime}\theta(\mu^{\prime}-m(\mu^{\prime}))(\mu^{\prime2}-m^{2}(\mu^{\prime}))^{3/2},
\end{equation}
where we have taken $N_{f}=N_{c}=3$. From Eq. (6) it can be
seen that the EOS of quasi-particle model depends on the model
parameters $a,c$. Moreover, the so-called vacuum pressure $B$ is
also a parameter which affects the EOS in the study of neutron-like
star \cite{LH,LH2,Yan}. In the following, we shall discuss the relation between parameters $a, c$ and $B$.

The choice of $B$ in quasi-particle model is complicated (depending
on the corresponding hadronic EOS \cite{APR} and quark EOS). The
different parameters $a, c$ in quasi-particle model lead to
different ranges of $B$. In order to see it clearly, let us to
define $\delta P(\mu) \equiv P _{quark}(\mu)-P_{hadron}(\mu)$ to
study the phase transition. Here $P_{quark}(\mu)$ and
$P_{hadron}(\mu)$ denote the pressure of quark matter and hadronic
matter, respectively. $P_{quark}(\mu)$ can be obtained using Eq.
(6). According to the EOS in Ref. \cite{APR}, $P_{hadron}=0$ when
$\mu<307 MeV$. A physical phase transition infers that $\delta
P(\mu)$ intersects with 0 once when  $\mu>307 MeV$, that is
$B>P_{quark}(0.307 GeV)|_{\mu\neq0}$, where
$P_{quark}(\mu)|_{\mu\neq0}$ stands for the second term in the
right-hand side of Eq. (6). However, the case is not always so
simple. We show $\delta P(\mu)$ in Fig. 1 with model parameters
$a=12/GeV$, $c=-0.05 GeV$. Just as is shown in Fig. 1, $\delta
P(\mu)$ is not even a monotonously increasing function. There are
two peaks $\delta P(\mu_{1})$ and $\delta P(\mu_{2})$,
$\mu_{1}<\mu_{2}$. We have two choices to get a valid result: (1)
$\delta P(\mu_{1})<0$, that is
$B>P_{quark}(\mu_{1})|_{\mu\neq0}-P_{hadron}(\mu_{1})$, i.e.,
$B^{1/4}=150 MeV$ in Fig. 1; (2) $\delta P(\mu_{2})>0$ and
$B>P_{quark}(0.307 GeV)$, that is $P_{quark}(0.307
GeV)|_{\mu\neq0}<B<P_{quark}(\mu_{2})|_{\mu\neq0}-P_{hadron}(\mu_{2})$,
i.e., $B^{1/4}=137 MeV$ in Fig. 1. It can be seen that some values
of $B$ (i.e., $B^{1/4}=140 MeV$) would lead to the result which is
not physical. It may be due to the inconsistency of QGP model and
constant vacuum pressure. At present we do not have a valid method
to deal with it, instead, when we use QGP model to study the
structure of the neutron-like star, we should pay attention to the
range of $B$. For some sets of model parameters $a, c$, the second
choice does not exist because
$P_{quark}(\mu_{2})|_{\mu\neq0}-P_{hadron}(\mu_{2})<P_{quark}(0.307
GeV)|_{\mu\neq0}$.

\begin{figure}
\begin{center}
\includegraphics[width=0.5\textwidth]{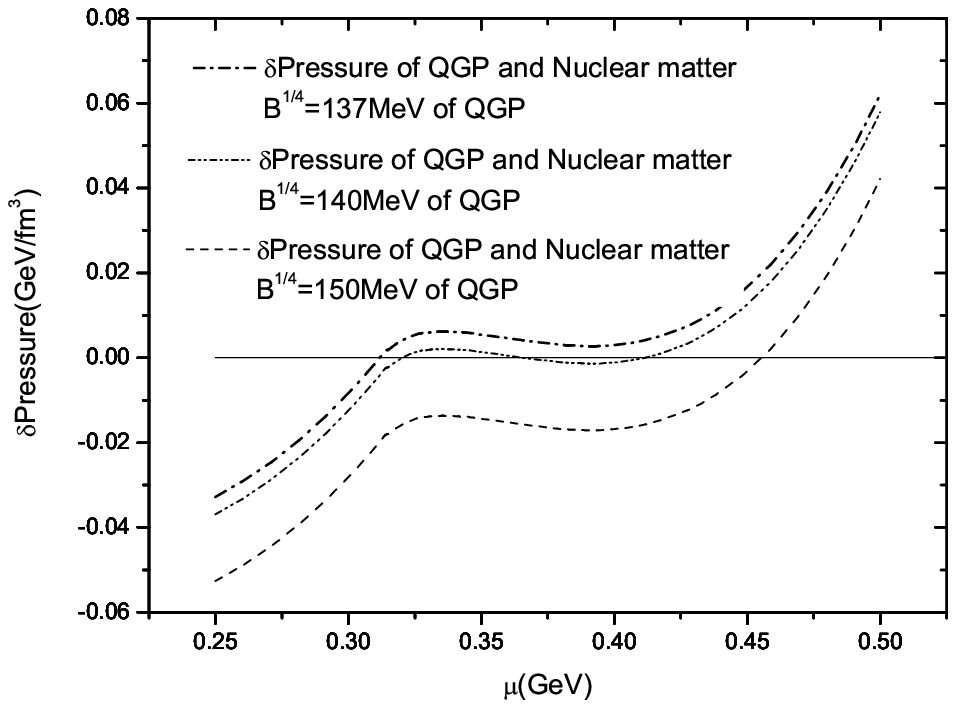}
\end{center}
\vspace{-0.8cm} Fig. 1. The $\delta Pressure$ of quark matter
described by quasi-particle model ($a=12/GeV, c=-0.05 GeV$) and hadronic
matter with different $B$
\end{figure}

The mass-radii relation of neutron-like star depends obviously on $B$. The mass-radii relation of the pure quark star
and the hybrid star (with $a=13/GeV, c=-0.1 GeV$) with different $B$
are plotted in Fig. 2 . From Fig. 2 it can be seen that for pure quark star
lager $B$ yields smaller mass and radius, while the case for hybrid
star is more complicated. We will discuss it in Part B of Section
III.

\begin{figure}
\begin{center}
\includegraphics[width=0.5\textwidth]{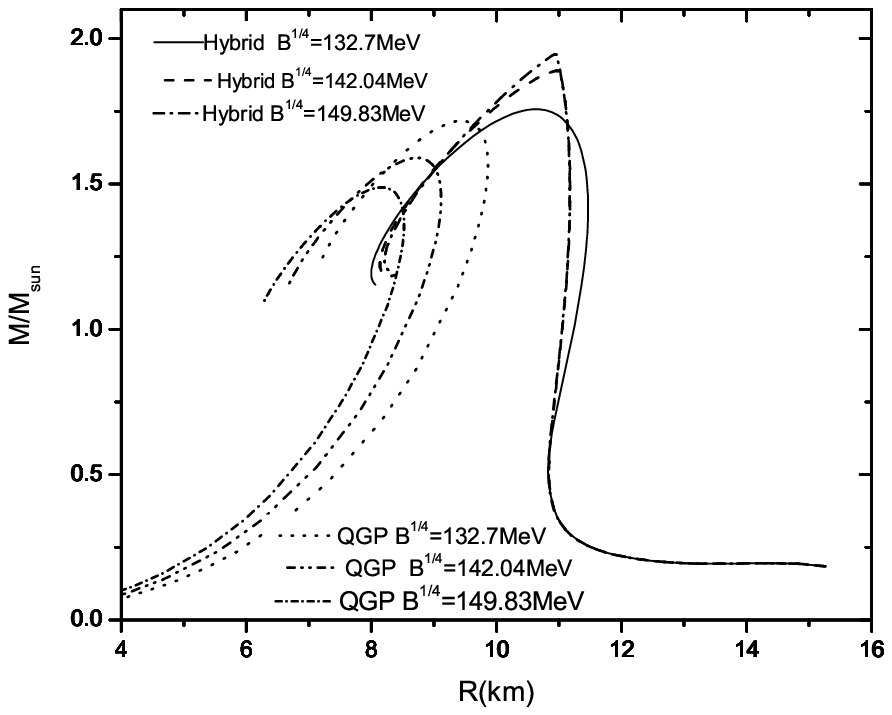}
\end{center}
\vspace{-0.8cm} Fig. 2. The mass-radii relations of hybrid star and
pure quark star with $a=13/GeV, c=-0.1 GeV$ and different $B$
\end{figure}

\section{The structure of neutron-like star}
Using the EOS, we can obtain the structure of a neutron-like star by
integrating the Tolman-Oppenheimer-Volkoff equation
\begin{eqnarray}
&&\frac{d P(r)}{dr}=-\frac{G(\varepsilon+P)(M+4\pi r^{3}P)}{r(r-2GM)},\\
&&\frac{d M(r)}{dr}=4\pi r^{2}\varepsilon.
\end{eqnarray}
The energy density reads
\begin{equation}
\varepsilon(\mu)=-P(\mu)+\mu\cdot\frac{\partial P}{\partial \mu}.
\end{equation}

In this paper, we consider the interior structure of a neutron-like
star composed, in whole or in part, of quark matter. The structure
of pure quark star is described totally by the quasi-particle model.
The hybrid star which is proposed by the authors of Refs.
\cite{K,K1,K2} has a quark core described by quasi-particle model
and a crust of nuclear matter described by the APR hadronic EOS
\cite{APR}.

\subsection{Pure Quark Star}
In this part, we use the recent observation of
$1.97\pm0.04$-solar-mass neutron-like star \cite{2solarmass} to
constrain the parameters $a, c$. In the previous section, we show
that the larger $B$ yields smaller mass, so we can choose the
smallest $B$ which leads to the valid result to get the maximal mass
for every set of $a, c$.

\begin{figure}
\begin{center}
\includegraphics[width=0.5\textwidth]{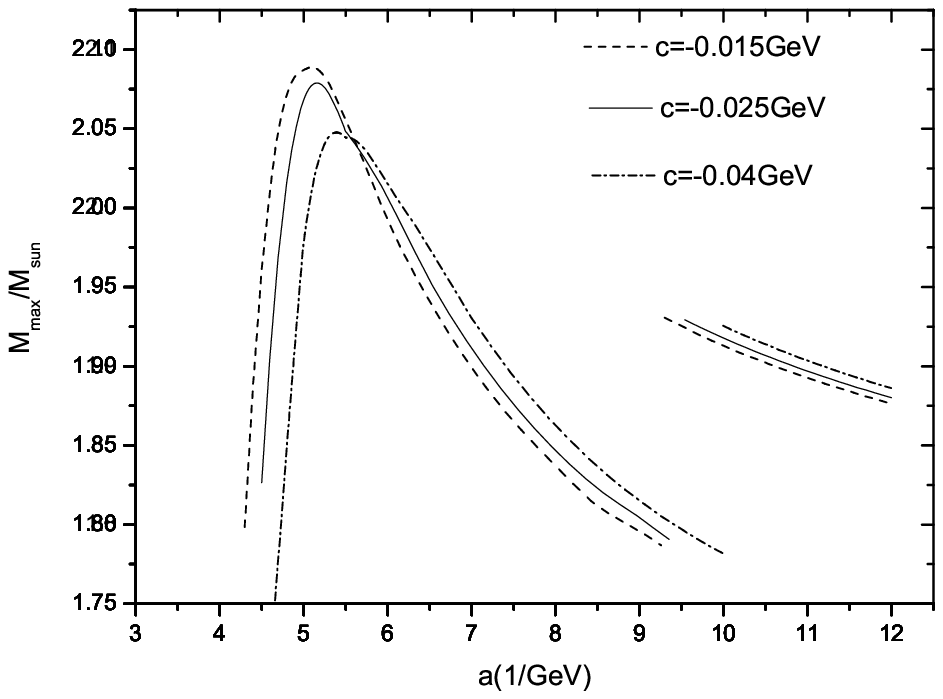}
\end{center}
\vspace{-0.8cm} Fig. 3.The maximal mass changes with the parameter
$a$ for different $c$.
\end{figure}

The relation between the mass and model parameters $a, c$ is shown
in Fig. 3. It can be seen clearly that for the same $c$, the mass for
different $a$ increases at first, then decreases later. This is
because at first the smallest $B \approx 0$, larger $a$ yields
larger mass. When $B$ dominates the maximal mass, larger $a$ leads
to larger $B$ which yields smaller mass. This two obvious regions
come from the different choices of $B$ which we discuss in Section
II. For different $c$, larger $c$ yields larger mass at first, then
the larger $c$ leads to smaller mass. The largest maximal mass to
$a$ increases with $c$ increasing.

With the mass related to the the parameters $a, c$, we can constrain
the parameters with the recent observation of
$1.97\pm0.04$-solar-mass neutron-like star \cite{2solarmass}. We
assume it is a pure quark star which can be described by
quasi-particle model.
\begin{figure}
\begin{center}
\includegraphics[width=0.5\textwidth]{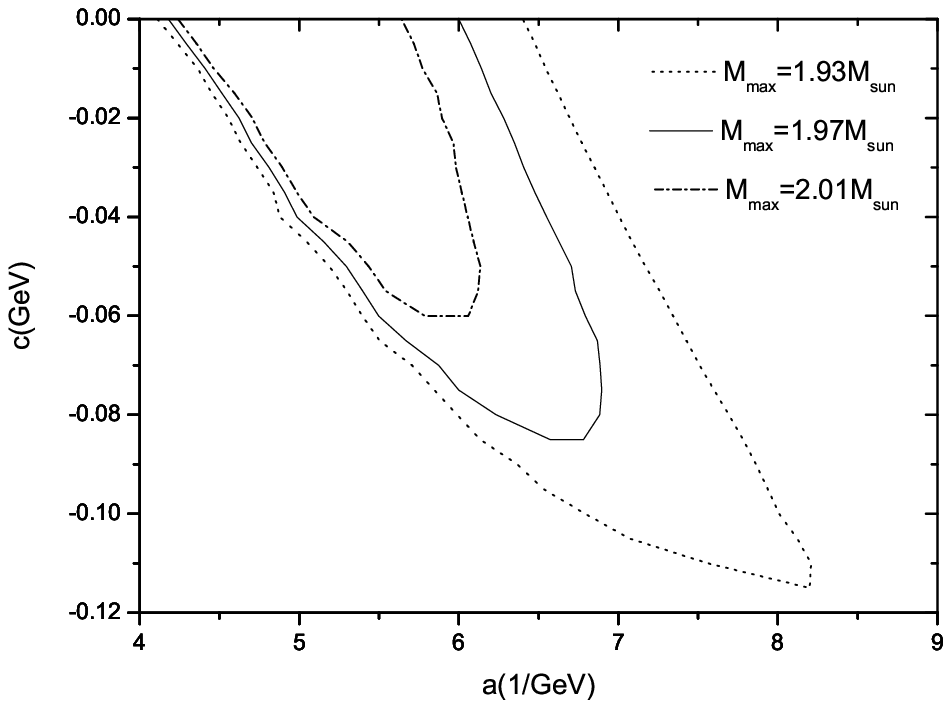}
\end{center}
\vspace{-0.8cm} Fig. 4. The contour of the parameters $a, c$
constrained by recent observation \cite{2solarmass}
\end{figure}

\begin{figure}
\begin{center}
\includegraphics[width=0.5\textwidth]{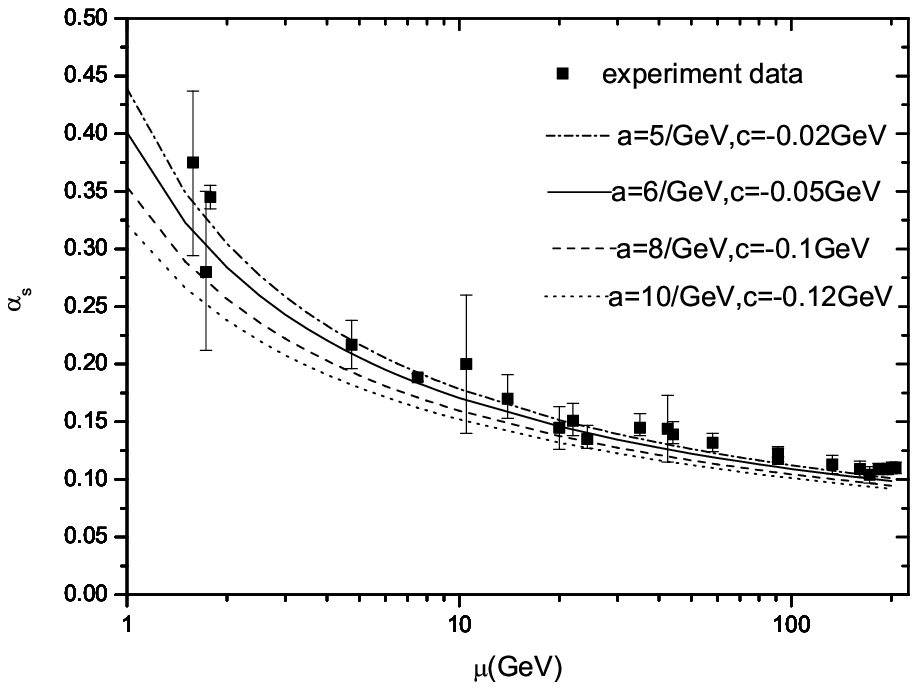}
\end{center}
\vspace{-0.8cm} Fig. 5. Comparison of the running coupling constant
$\alpha_s$ calculated using different values of $a, c$ with the
experimental data. The experimental data are taken from table 1 of
Ref. \cite{exp}.
\end{figure}

\begin{figure}
\begin{center}
\includegraphics[width=0.5\textwidth]{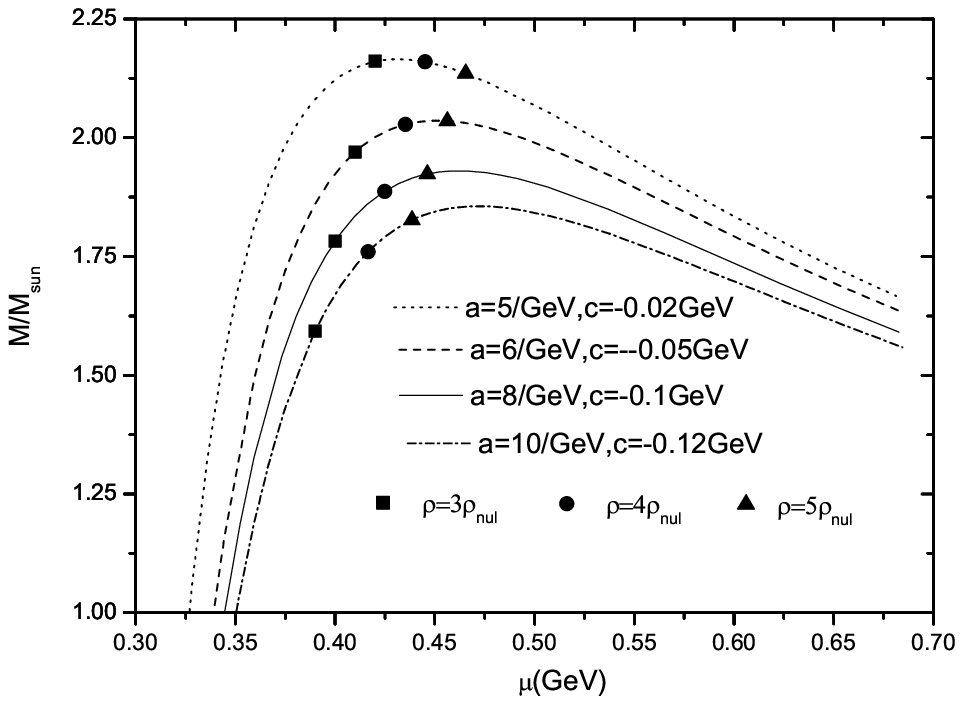}
\end{center}
\vspace{-0.8cm} Fig. 6. Mass of different central potential with
different parameters $a,~c$, where $\rho_{nul}=2.8 \times
10^{14}g/cm^{3}$. $B$ is chosen to be smallest which is physical to
obtain the maximal mass.
\end{figure}
The range of $a, c$ is illustrated in Fig. 4. The physical meaning
of the parameters $a, c$ is not obvious. Here, we try to understand
it from the point of view of the running coupling constant
$\alpha_s$. The comparison of the experimental data on $\alpha_s$
and the results of $\alpha_s$ calculated with some representative
values of $a, c$ are shown in Fig. 5. Here, one point should be
clarified. In the usual expression of the running coupling constant
$\alpha_s(Q^2)$, $\alpha_s$ is a function of the momentum scale $Q$.
In our work, in order to understand the physical meaning of
$\alpha_s$, the chemical potential $\mu$ is taken as an energy scale
similar to the momentum scale $Q$ in the expression of $\alpha_s$. From Fig. 5 and Fig. 6 (see below), it can be
seen that the parameters which yield $1.97\pm0.04 M_{\bigodot}$ are
also consistent with the experimental data on $\alpha_s$ \cite{exp}.
Therefore, we conclude that quasi-particle model with proper
parameters is suitable to describe quark matter. 

If we assume the
central density $\mu$ of quark star to be $0.45 GeV$, from Fig. 6 we
deduce it is reasonable, then the range of $\alpha_{s}$ is from
$0.659 \sim 1.135$. By calculation we can also find that the range
of $\mu_{0}$ is $0.21 GeV \sim 0.327 GeV$. In Ref. \cite{Halasz},
based on a universal argument, it is pointed out that the existence
of some singularity at the point $\mu=\mu_0$ and $T=0$ is a robust
and model-independent prediction. The numerical value of critical
chemical potential in pure QCD (i.e., with electromagnetic
interactions being switched off) is estimated to be $\frac{m_{N}-16
~\rm{MeV}}{N_{c}}=307 ~\rm{MeV}$ (where $m_{N}$ is the nucleon mass
and $N_{c}=3$ is the number of colors). Here, in order to be self-contained, we shall give an explanation for how this result
is derived. According to the argument of Ref. \cite{Halasz}, the
critical baryon chemical potential $\mu_{B0}={\rm
min}_\alpha(E_\alpha/N_\alpha)$, where $\alpha$ stands for a generic
quantum state of the system and $E_\alpha$ and $N_\alpha$ is the
energy and baryon number of this state, respectively. The energy per
baryon, $E/N$, can also be written as $m_N-(N m_N-E)/N$, where $m_N$
is the nucleon mass. Therefore, the state which minimizes $E/N$ is
that for which the binding energy per nucleon, $\epsilon=(N
m_N-E)/N$, is a maximum. Empirically, we know that this state is a
single iron nucleus at rest with $N=A=56$ and $\epsilon \approx 8
~\rm{MeV}$. However, in QCD without electromagnetism the binding
energy per nucleon increases with $N=A$. This is the consequence of
the saturation of nuclear forces and can be seen from the Weizsacker
formula. Without electromagnetism, only the bulk and surface energy
terms are significant for large $A$:
\begin{eqnarray}
\epsilon(A)\equiv \frac{A m_N-m_A}{A}\approx a_1-a_2 A^{-1/3} \nonumber
\end{eqnarray}
with $a_1 \approx 16 ~\rm{MeV}$, $a_2 \approx 18 ~\rm{MeV}$ \cite{FW}. As $A \rightarrow \infty$, $\epsilon$ saturates at the value $a_1$. This corresponds to the binding energy per nucleon in a macroscopically large sample of nuclear matter as defined by Fetter and Walecka in \cite{FW}. Therefore, in QCD the critical baryon chemical potential $\mu_{B0}\approx m_N-16 ~\rm{MeV}$ and the corresponding quark chemical potential
$\mu_0$ is
$\frac{m_{N}-16~ \rm{MeV}}{N_{c}}=307~ \rm{MeV}$.
The range of parameters $a, c$ constrained by the observation of
$1.97\pm0.04$-solar-mass neutron-like star \cite{2solarmass} is
consistent with it.

\subsection{Hybrid Star}

The hybrid star was inspired by the idea of the phase transition from
quark matter to nuclear matter \cite{Hybrid,K,K1,K2}. The phase
transition in this paper is modeled by a simple Maxwell's
construction.

\begin{figure}
\begin{center}
\includegraphics[width=0.5\textwidth]{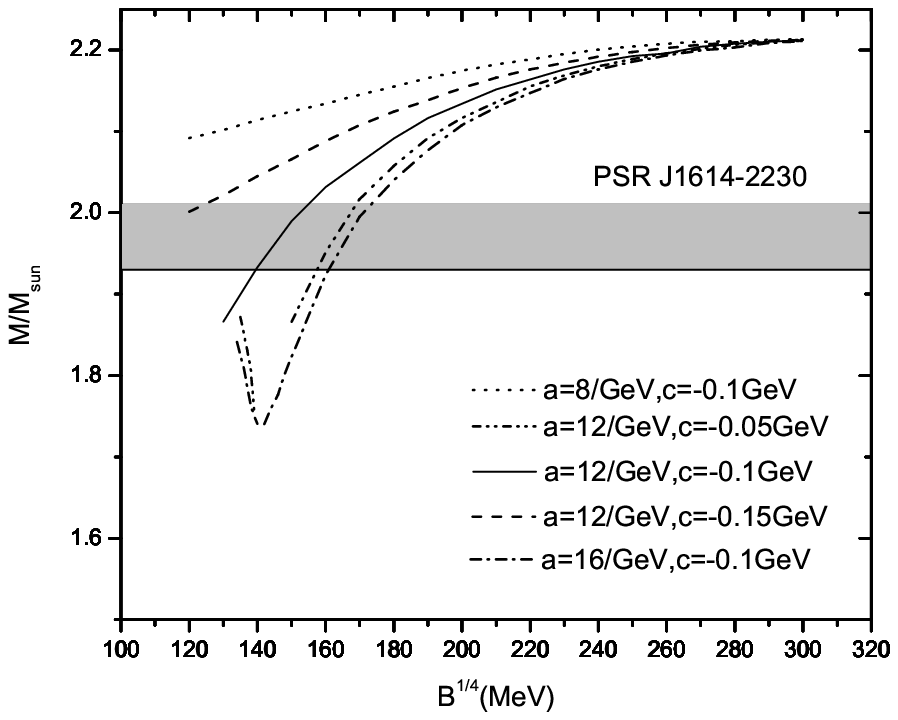}
\end{center}
\vspace{-0.8cm} Fig. 7. The maximal mass changes with the parameter
$B$ with different $a,~c$.
\end{figure}

In Fig. 7 we plot the relation of the maximal mass to the parameter
$B$. From Fig. 7 it can be seen that in the large $B$ region, the
maximal mass tends to $2.2 M_{\bigodot}$. This means that for large $B$,
the hybrid star
tends to a star which is composed totally by hadronic matter. The
maximal mass for small $B$ depends on the parameters in
quasi-particle model. For some parameters (e.g., $a=12/GeV, c=-0.1
GeV$), the maximal mass increases with $B$ increasing, while for
other parameters (e.g., $a=12/GeV,~ c=-0.05 GeV$), the maximal mass
decreases with $B$ increasing at the beginning, then it increases
with $B$ increasing. This means that for the latter parameters, the
behavior at small $B$ tends to a pure quark star. The discontinuous
curves with parameters $a=12/GeV,~c=-0.05 GeV$ and $a=16/GeV,~c=-0.1
GeV$ shown in Fig. 7 is caused by the valid ranges of $B$. We can
compare the result with the same plot published in
Ref. \cite{MIT2011} which has a quark core described by MIT bag model
and hadronic crust described by TM1 RMF EOS and NL3 RMF EOS.
However, the running coupling constant in the center of the star in
Fig. 7 is narrow. In Fig. 8, we plot the maximal mass to $B$ with
different central coupling constant. The plot is similar with the
result in Ref. \cite{MIT2011}. In Ref. \cite{MIT2011}, the authors
stressed that different hadronic model could lead to the differences of
the result. In our work, we use APR EOS to describe the hadronic crust.
\begin{figure}
\begin{center}
\includegraphics[width=0.5\textwidth]{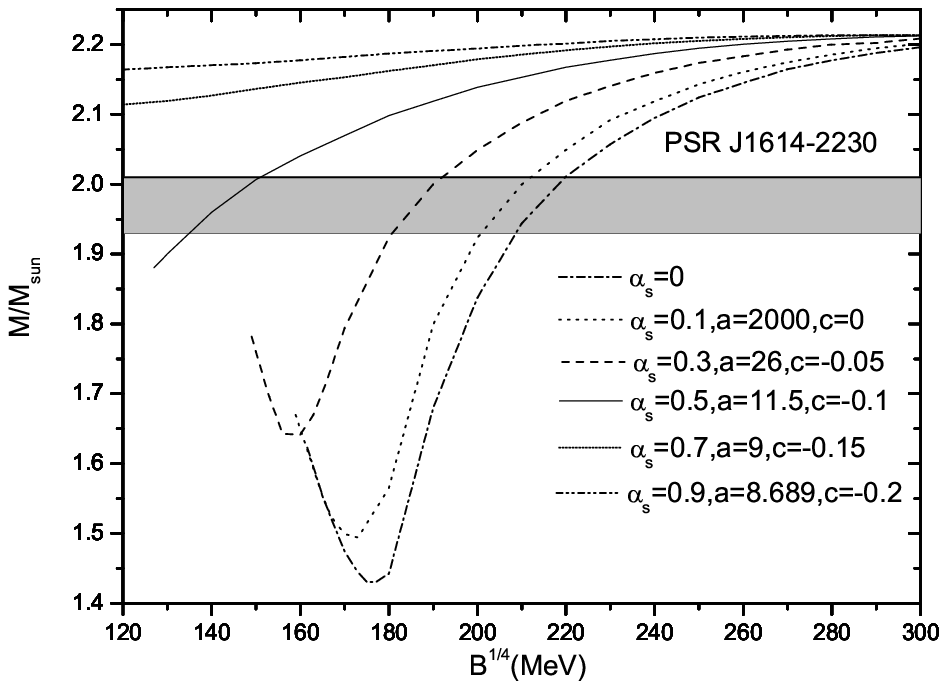}
\end{center}
\vspace{-0.8cm} Fig. 8. The maximal mass changes with the parameter
$B$ with different $\alpha_{s}$.
\end{figure}

From Fig. 9 and Fig. 11 , a smaller $B$ will narrow the region of
$a$-$c$ space. However, for hybrid star we cannot give a rigid range
of $a$ and $c$. It can also be figured out that smaller $a$ and $c$
will yield larger maximal mass. According to Eq. (3), 
larger running coupling constant leads to larger mass, which is
consistent with the $1.97\pm0.04$ solar-mass neutron-like star.

\begin{figure}
\begin{center}
\includegraphics[width=0.5\textwidth]{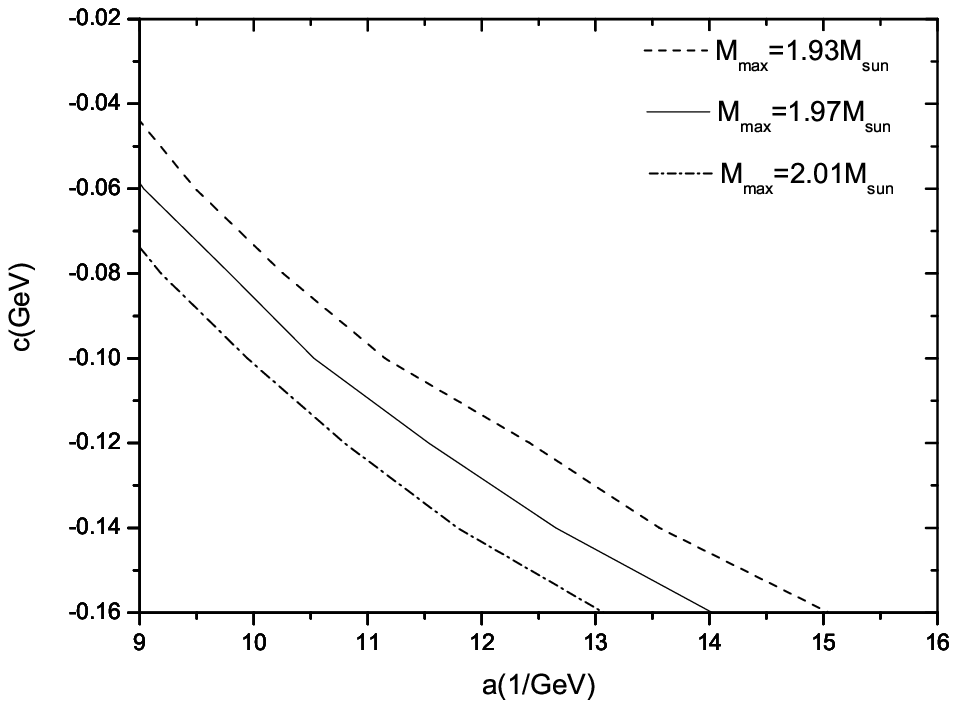}
\end{center}
\vspace{-0.8cm} Fig. 9. The contour picture of parameters in
$B^{1/4}=130.58 MeV$
\end{figure}
\begin{figure}
\begin{center}
\includegraphics[width=0.5\textwidth]{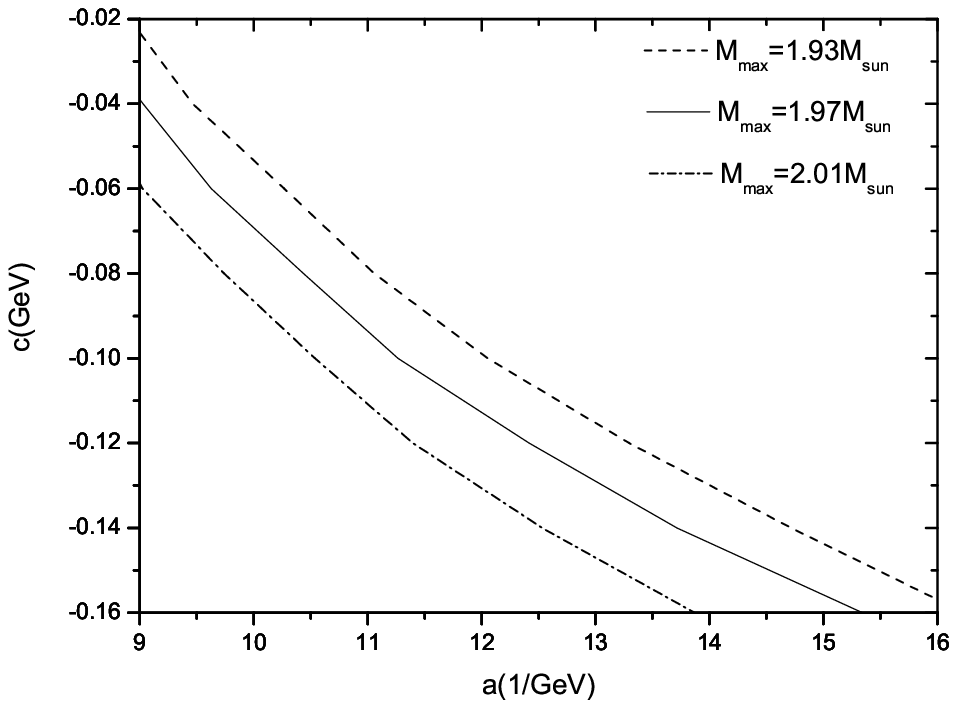}
\end{center}
\vspace{-0.8cm} Fig. 10. The contour picture of parameters in
$B^{1/4}=140.32 MeV$
\end{figure}
\begin{figure}
\begin{center}
\includegraphics[width=0.5\textwidth]{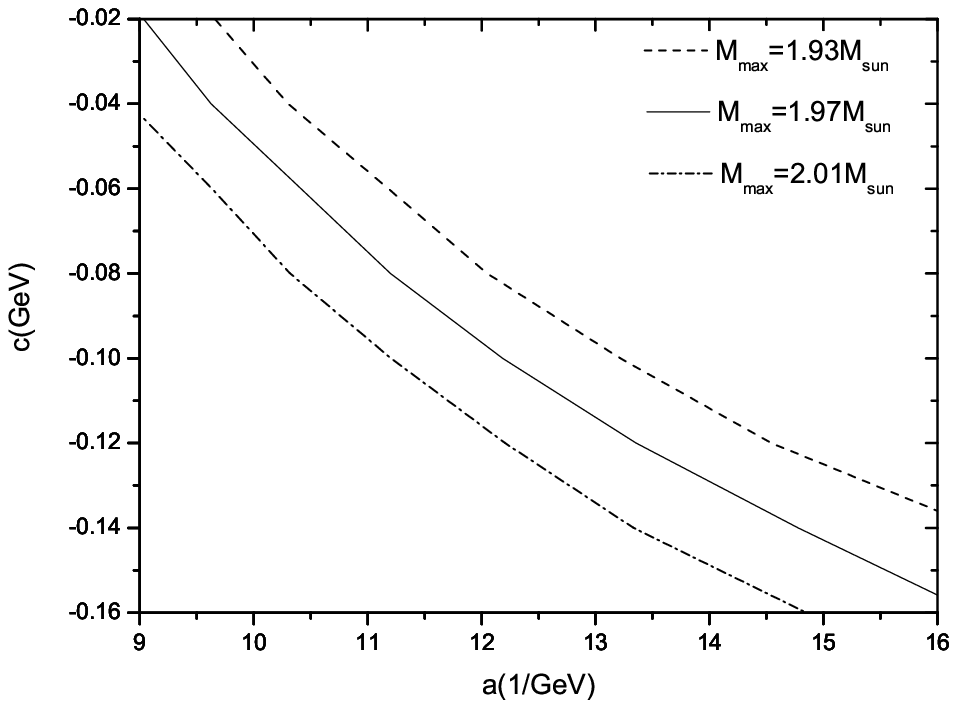}
\end{center}
\vspace{-0.8cm} Fig. 11. The contour picture of parameters in
$B^{1/4}=148.37 MeV$
\end{figure}

\section{Summary}

The parameters of quasi-particle model at zero temperature and
finite chemical potential can be constrained by recent observation
\cite{2solarmass}. We consider the structure of a neutron star which
is composed, in whole or in part, of quark matter. In this paper
quark matter is
described by quasi-particle model incorporating a bag constant $B$.

We discuss the range of $B$ and the influences of $B$ on both quark
and hybrid star. First, we assume the neutron-like star is a pure quark star. Larger $B$ yields smaller mass. We can constrain the parameters
$a, c$ by recent observation \cite{2solarmass}. The range of the running
coupling constant $\alpha_{s}$ with central density $\mu = 0.45 GeV$ of pure
quark star is $0.659 \sim 1.135$ (this clearly shows that even at such a high density the coupling constant is still large, so perturbation theory cannot be used). We can also find that the range of
$\mu_{0}$ is $0.21 GeV \sim 0.3 27 GeV$, which is consistent with the
result given in Ref. \cite{Halasz}. Second, we assume the neutron-like star to be a hybrid star which has a
quark core and a hadronic crust. The parameters of quasi-particle
model and the vacuum pressure $B$ affect the structure of hybrid
star. In the large $B$ region, the maximal mass tends to $2.2
M_{\bigodot}$. The maximal mass for small $B$ depends on the
parameters of quasi-particle model. For some parameters (e.g.,
$a=12/GeV, c=-0.1 GeV$), the maximal mass increases with $B$
increasing, while for other parameters (e.g.,
$a=12/GeV, c=-0.05 GeV$), the maximal mass decreases with $B$
increasing at the beginning, then it increases with $B$ increasing.
Larger $a$ and $c$ lead to smaller mass of hybrid star.  We conclude that either a pure quark star and a hybrid star with a quark core described by quasi-particle model can be consistent with the recent observation
of PSR J1614-2230. The parameters in quasi-particle model can be
narrowed by the observation.

\bigskip

This work is supported in part by the National Natural Science
Foundation of China (under Grant Nos 10775069, 10935001 and
11075075) and the Research Fund for the Doctoral Program of Higher
Education (under Grant No 200802840009)

\end{document}